\documentclass[a4paper,10pt]{article}
\usepackage{subfigure}
\usepackage{graphicx}
\usepackage{tikz}
\usepackage{amsfonts}
\usepackage{booktabs}
\newcommand{\ra}[1]{\renewcommand{\arraystretch}{#1}}
\newcommand{\unit}[1]{\ensuremath{\,\mathrm{#1}}}
\newcommand{\const}[0]{\ensuremath{\mathrm{const}}}

\newcommand{\rmd}[0]{\ensuremath{\mathrm{d}}}
\newcommand{\rme}[0]{\ensuremath{\mathrm{e}}}
\newcommand{\rmi}[0]{\ensuremath{\mathrm{i}}}
\begin{document}

\title{Metropolis-Hastings thermal state sampling\\
for numerical simulations\\
of Bose-Einstein condensates}

\author{Pjotrs Gri\v{s}ins$^{1,*}$ and Igor E Mazets$^{1,2}$\vspace{3mm}\\
\normalsize $^1$ Vienna~Center~for~Quantum~Science~and~Technology,\\
\normalsize  Atominstitut~TU~Wien, 1020~Vienna, Austria \\
\normalsize $^2$ Ioffe~Physico-Technical~Institute, Russian~Academy~of~Sciences,\\
\normalsize 194021~St.~Petersburg, Russia \\
\normalsize $^*$ \texttt{pgrisins@ati.ac.at}}

\maketitle

\begin{abstract}
We demonstrate the application of the Metropolis-Hastings algorithm to sampling of classical thermal states of one-dimensional Bose-Einstein quasicondensates in the classical fields approximation, both in untrapped and harmonically trapped case. The presented algorithm can be easily generalized to higher dimensions and arbitrary trap geometry. For truncated Wigner simulations the quantum noise can be added with conventional methods (half a quantum of energy in every mode). The advantage of the presented method over the usual analytical and stochastic ones lies in its ability to sample not only from canonical and grand canonical distributions, but also from the generalized Gibbs ensemble, which can help to shed new light on thermodynamics of integrable systems.

\end{abstract}

\section{Introduction}

The recent advances in experimental methods allowed precise control and manipulation of ultracold atoms in various trap \cite{Berrada2013a,Desbuquois2012,Hung2011} and optical lattice geometries \cite{Bloch2012,Reinhard2013,Zhang2012}, including gases at temperatures much lower than the degeneracy temperature. 


The effective field theory of a cold gas of neutral bosonic atoms with short-range repulsive interactions is given by the second quantized Hamiltonian (in the following we deal explicitly with a one-dimensional (1D) case, where quasicondensation takes place instead of true condensation \cite{Haldane1981})
\begin{equation}
	\hat H = \hat H_0 + \hat H_{int},
\end{equation}
\begin{equation}
   \hat H_0 = \int \rmd z\; \hat\psi^\dag(z) \left[ -\frac{\hbar^2}{2m} \frac{\partial^2}{\partial z^2} + V(z) 
		\right] \hat\psi(z),
\end{equation}
\begin{equation}
	\hat H_{int} = \frac{g \hbar}{2} \int \rmd z\; \hat\psi^\dag(z)\, \hat\psi^\dag(z)\, \hat\psi(z)\, \hat\psi(z),
\end{equation}
where $\hat H_0$ and $\hat H_{int}$ are respectively the free-particle and interaction Hamiltonians, $\hat\psi(z)$ is the field operator, which annihilates a particle at position $z$, $m$ is the atomic mass, $V(z)$ is the external trap potential and $g$ is the effective interaction strength, given in the experimentally relevant case of a harmonic transversal confinement with trapping frequency $\omega_r$ by $g = 2 \omega_r a_s$, with $a_s$ being the s-wave scattering length.


The usual experimental setups deal with thousands of atoms \cite{Berrada2013a}, so the quantum dynamics can be numerically simulated only using various approximations. The one approximation especially suited for studies of weakly interacting cold atomic gases is the classical field approximation, where we replace the quantum field operator of the effective field theory $\hat\psi(z)$ by a classical field $\psi(z)$ \cite{Blakie2008}. This approach is valid for low temperatures, where we have a range of macroscopically occupied modes $\langle|\hat\psi_n|^2\rangle \gg 1$; the modes $\hat\psi_n$ are taken to be the eigenfunctions of the one-body non-interacting Hamiltonian $\hat H_0$. The evolution of this redefined classical order parameter $\psi(z)$ is then governed by the celebrated Gross-Pitaevskii equation (GPE) \cite{Dalfovo1999}.

In experiments with cold atomic gases the system is usually prepared in thermal equilibrium, before a quench or another manipulation is applied, therefore the numerical methods for sampling the thermal initial condition $\psi_0(z)$ are of great importance. The quantum correction for the classical thermal state of a weakly interacting system can be introduced using the so called truncated Wigner approximation (TWA), where zero-point quantum oscillations are taken into account in the initial state only, but the subsequent evolution is classical \cite{Polkovnikov2010}.

Conventional methods of initial state sampling include analytical ones \cite{Stimming2011,Grisins2011}, where the gas is initialized with a Bose-Einstein distribution of Bogoliubov quasiparticles with random phases, as well as stochastic ones \cite{Cockburn2011,Duine2001}, where the thermal state is achieved during imaginary time GPE evolution with Langevin noise.

In the present paper we propose another way of sampling the initial distribution, namely using the Metropolis-Hastings algorithm. We believe that in some cases it might be preferable over the analytical methods, as it does not use Bogoliubov-type approximations, and may be used to sample states out of a generalized Gibbs ensemble, which is impossible with existing stochastic realizations.

\section{Metropolis-Hastings algorithm}

The Metropolis-Hastings algorithm is a Markov chain Monte Carlo method for sampling a probability distribution, especially suited for systems with many degrees of freedom \cite{Metropolis1953}. For a broad overview of quantum and classical Monte Carlo methods, including the Metropolis-Hastings algorithm, see \cite{Newman1999,Anderson2007} and references therein.



In the present paper we demonstrate the implementation of the Metropolis-Hastings method for 1D Bose-Einstein quasicondensate without confinement (implying periodic boundary conditions) as well as for the experimentally relevant case of a harmonic longitudinal confinement. The method can be easily generalized to higher dimensions and other trap geometries.

This method has been already applied to classical simulations of cold Bose gases \cite{Bienias2011}, but it has not been explicitly formulated as a step-by-step algorithm. In the present paper we systematically study the convergence properties of this method and outline its application to sampling the generalized Gibbs ensemble (GGE).

In our particular realization the algorithm reads as follows:
\begin{enumerate}
\item Initialization: 

\begin{enumerate}
\item Choose an initial order parameter $\psi_0(z)$. Specific choices of $\psi_0(z)$ will be discussed in the following section.
\item Calculate the reduced entropy $S_0 = -\beta\left(\langle \psi_0 | \hat H | \psi_0 \rangle - \mu \langle \psi_0 | \hat \mathcal{N} | \psi_0 \rangle\right)$, where $\beta$ is the inverse temperature, $\mu$ is the chemical potential (both $\beta$ and $\mu$ are fixed external parameters), and $\hat \mathcal{N} = \int \hat \psi^\dag(z)\,\hat \psi(z)\,\rmd z$ is the particle number operator. Note that the free energy does not enter the expression for $S_0$, meaning that the zero-level of the latter is not defined. This is justified by the fact that we are interested only in differences of the reduced entropy, and not its absolute value.
\end{enumerate}
\item For each iteration $N \in [1, N_{\max}]$:

\begin{enumerate}
\item Generate a candidate field $\psi_N(z)$ by varying the energy. This variation of energy can be achieved by adding either a density phonon
\begin{equation}
	\psi_N(z) = \psi_{N-1}(z) \sqrt{ 1 + c_1 v_r \sin(k_r z + \phi_r)},
\end{equation}
or a phase phonon
\begin{equation}\label{e:1}
	\psi_N(z) = \psi_{N-1}(z) \exp\left[\rmi\; c_2 v_r \sin(k_r z + \phi_r)\right],
\end{equation}
to the field from the previous iteration (`the seeding field'). Whether to choose the one or the other is decided at random (by a `coin toss'). The meaning and values of the parameters are summarized in table~\ref{t:1}.
\item\label{i:1} Vary the particle number
\begin{equation}\label{e:2}
	\psi_N(z) = (1 + c_3 u_r)\psi_N(z).
\end{equation}
\item Calculate the reduced entropy of the candidate field
\begin{equation}
S_N = -\beta\left(\langle \psi_N | \hat H | \psi_N \rangle - \mu \langle \psi_N | \hat \mathcal{N} | \psi_N \rangle\right)
\end{equation}
\item Calculate the acceptance ratio $a = \frac{p_N}{p_{N-1}} = \frac{\rme^{S_N}}{\rme^{S_{N-1}}} = \rme^{S_N-S_{N-1}}$, where $p_N = \frac{1}{Z} \rme^{S_N}$ is the Boltzmann probability to find the field in the state $\psi_N(z)$. The main advantage of the Metropolis-Hastings algorithm lies the fact that we have to evaluate only the ratio of probabilities, in this way avoiding to calculate the partition function $Z$, which is practically impossible for interacting systems with many degrees of freedom. Then we check the value of $a$:
\begin{enumerate}
\item If $a\geq 1$, then the candidate state is more probable than the seeding state, so we keep the former.
\item If $a<1$, we pick a uniform random number $r\in[0,1]$. If $r\le a$, the candidate state is accepted; but if $r>a$, the candidate state is discarded and the seeding state is kept for the next iteration $\psi_N(z) := \psi_{N-1}(z)$.
\end{enumerate}
\item Proceed to the next iteration.
\end{enumerate}
\end{enumerate}

\begin{table}
\caption{Numerical parameters of the Metropolis-Hastings algorithm.}
\centering
\ra{1.2}
\begin{tabular}{@{}lp{9cm}@{}}
\toprule
Parameter & Description\\
\midrule
$v_r, u_r$ & Real random numbers, distributed normally with zero mean and unit variance. \\ 
$c_1$, $c_2$, $c_3$  & Numerical constants governing the rate of convergence to the equilibrium state. In the presented results they have been empirically chosen to be $c_1 = 4 (n_0)^{-1/2}, c_2 = 0.1$ and $c_3 = 0.001$, where $n_0 = \max |\psi_0|^2$ is the maximal initial density. This particular choice provided typical values of the acceptance ratio in each iteration $a\in[0.4, 0.6]$, which gave the fastest convergence to equilibrium. It was numerically checked that different choices of those constants did not affect the resulting state, only the rate of convergence.\\
$\phi_r$ & Random phase $\phi_r\in[\,0,2\pi)$ picked from the uniform distribution. \\
$k_r$ & Random wave number picked from the set $\{\pm \delta k, \pm 2\delta k, \ldots, \pm k_{\max}\}$, where $\delta k = 2\pi/L$, $L$ is the length of the simulated region and $k_{\max}$ is the cutoff wave number. It was numerically checked that the results do not depend on this cutoff as long as it is larger than the inverse healing length $\xi^{-1} = \sqrt{m g \bar n/ \hbar}$, where $\bar n$ is the mean density. So we present results where $k_{\max} = \mathfrak{N}_z\,\delta k /2$ is the maximal allowed wave number on a lattice of $\mathfrak{N}_z$ sampling points.\\
\bottomrule
\end{tabular}
\label{t:1}
\end{table}

As a result we have a Markov chain of states $\psi_N(z)$, $N\in[0,N_{\max}]$, which can be used as thermal initial states for classical fields simulations. It is important to throw away the states obtained at early iterations (so called `burn-in' period), where the thermal state is not yet achieved. Neighbouring states $\psi_N$ and $\psi_{N+1}$ are usually highly correlated (as they differ by only one phonon and a small particle number variation), so it is necessary to throw away majority of the results, picking only one state out of $N_c$, where $N_c$ is calculated from the iteration-to-iteration autocorrelation length. We will return to these two issues in the results section.

Straightforward generalizations of the algorithm are easily conceivable:
\begin{enumerate}
\item Arbitrary trap geometry, as we can freely modify the trapping potential $V$ in the total Hamiltonian $\hat H$. In general the phonons in Eqs.~\ref{e:1} and~\ref{e:2} can be modified to be the eigenfunctions of the trapping potential (e.g. in the case of harmonic confinement $V(z) \propto z^2$ we can take Hermite functions instead of sine-waves). But in practice using potential-specific eigenfunctions instead of plane waves did not give any speed-up to the achievement of the steady state, so the algorithm can be used without this modification. 
\item Any number of dimensions. This requires representing the order parameter as a scalar field on many-dimensional space $\psi(\vec z)$, the phonons (Eqs.~\ref{e:1} and~\ref{e:2}) being modified accordingly as $\sin(\vec{k_r}\vec{z} + \phi_r)$.
\item Canonical state sampling. Reduced entropy becomes $S_N = -\beta \langle \psi_N | \hat H | \psi_N \rangle$, and we have to omit the \ref{i:1} stage of the algorithm to make sure the particle number does not change.
\item Generalized Gibbs ensemble sampling. Reduced entropy now reads
\begin{equation}
S_N = -\beta\left(\langle \psi_N | \hat H | \psi_N \rangle - 
\mu \langle \psi_N | \hat \mathcal{N} | \psi_N \rangle - \sum_i \mu_i \langle \psi_N | \hat I_i | \psi_N \rangle\right),
\end{equation}
where $\hat I_i$ are the local conserved charges (integrals of motion) of the system, in addition to the energy $\langle\hat H\rangle$ and the particle number $\langle\hat \mathcal{N}\rangle$, and $\mu_i$ are generalized potentials. For instance, in case of 1D GPE there exists an infinite number of local conserved charges, which can be explicitly calculated using Zakharov-Shabat construction \cite{Zakharov1973a}. We regard this possibility of GGE sampling as the primary advantage of the presented algorithm. In fact, simulation of the GGE requires only redefinition of the Hamiltonian to $\hat H' = \hat H - \frac{1}{\beta}\sum_i \mu_i \hat I_i$, to which the previously described algorithm can be applied without further modification. We reserve the detailed analysis of this case for a separate publication. 
\end{enumerate}

\section{Results}

\begin{figure}
 \centering
 \begin{tabular}{cc}
  \includegraphics[width=0.46\textwidth]{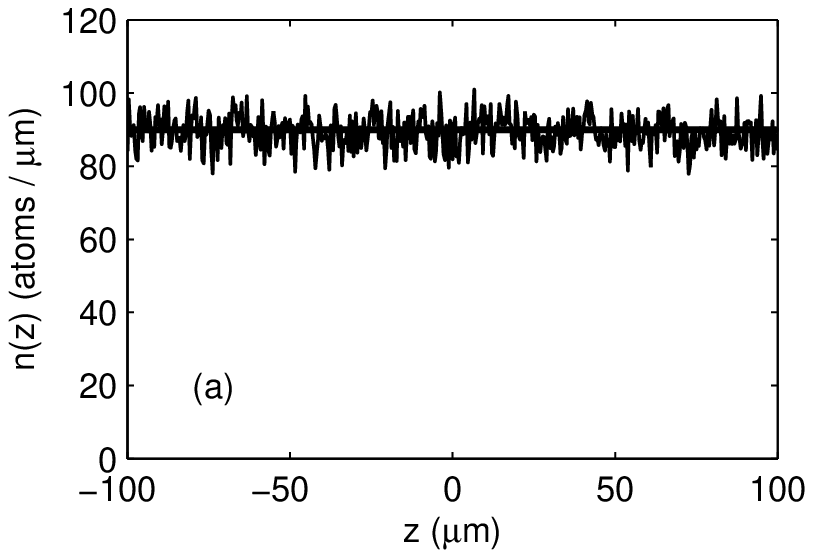}
    & 
  \includegraphics[width=0.46\textwidth]{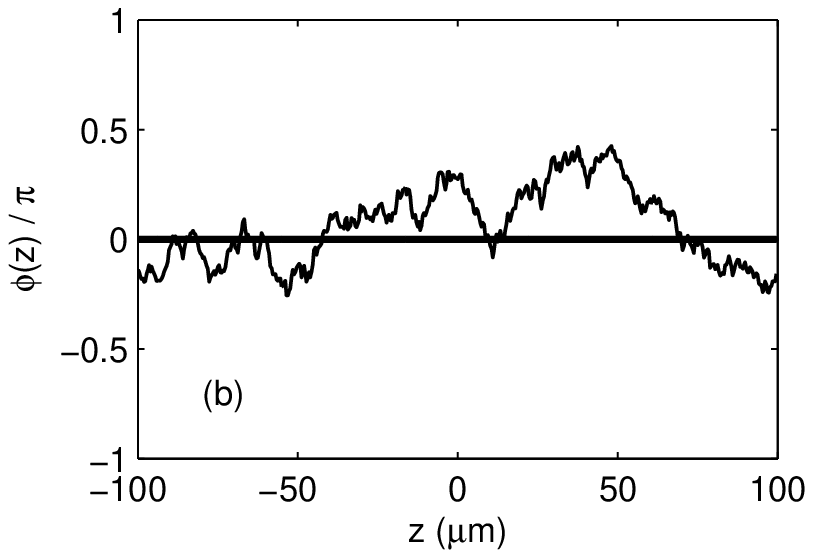}
    \\ 
  \includegraphics[width=0.46\textwidth]{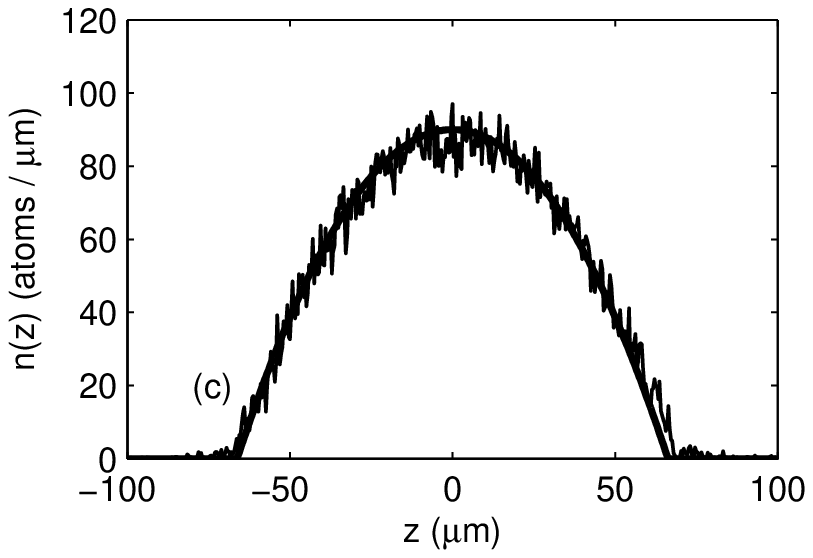}
    & 
  \includegraphics[width=0.46\textwidth]{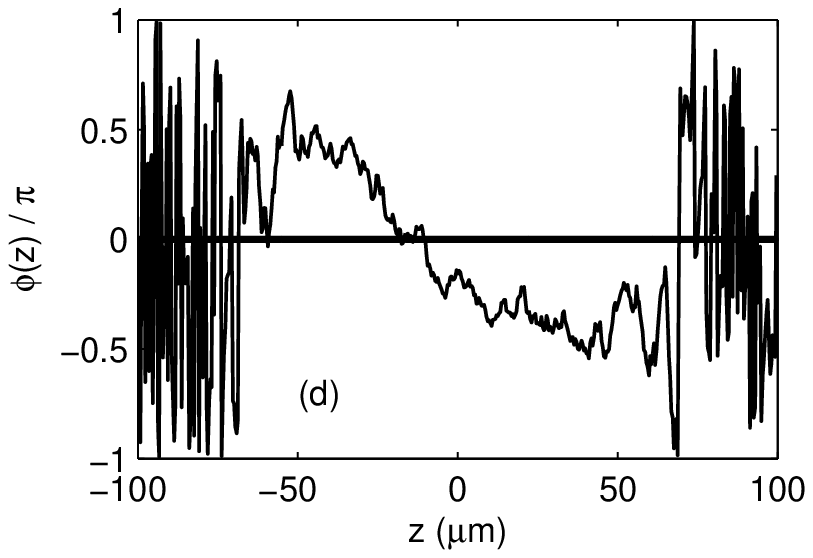}
 \end{tabular} 
 \caption{Typical examples of the grand canonical thermal state with the temperature $T=10\unit{nK}$ of the interacting 1D BEC, achieved after $N_{\max} = 10^5$ Metropolis-Hastings iterations in the untrapped system with periodic boundary conditions (a, b; thin zigzag line) and harmonically trapped case (c, d; thin zigzag line). Quasicondensate local densities $n(z)$ (a, c), measured in atoms per micrometer, and phases $\phi(z)$ (b, d), measured in units of $\pi$, as a function of the longitudinal direction $z$ in micrometers. Thick horizontal (a, b, d) and parabolic (c) lines represent the initial conditions, which in the case of the untrapped system (a, b) were taken to be the ground state of the non-interacting gas ($n_0(z) = n_0$, $\phi_0(z) = 0$), and in the case of the harmonic confinement (c, d) as a Thomas-Fermi parabolic density profile with constant zero phase. Extensive fluctuations of the phase at the edges of (d) are due to the fact that the density there is close to zero, and the phase can take arbitrary values. Physical parameters of the simulations are summarized in table~\ref{t:2}.}
\label{fig:profiles}
\end{figure}

In the following we demonstrate the application of the algorithm to generate a grand canonical thermal state for  an untrapped gas of neutral $^{87}$Rb atoms and an experimentally relevant case of the same gas in a harmonic confinement. The parameters of the simulation are summarized in table~\ref{t:2}.

\begin{table}
\caption{Simulation parameters of the systems presented in the results section.}
\centering
\ra{1.2}
\begin{tabular}{@{}lp{8cm}@{}}
\toprule
Parameter & Description\\
\midrule
$m = 87 \cdot 1.67 \cdot 10^{-24}\unit{g}$ & atomic mass of $^{87}$Rb atoms \\
$a_s = 5.3 \cdot 10^{-7}$ cm & s-wave scattering length \\ 
$T = 10, 60$ or $120$ nK & temperature \\
$\omega_r = 2\pi \cdot 2000 \unit{s^{-1}}$ & transversal trapping frequency \\
$n_0 = 90\unit{atoms/\mu m}$ & maximal initial linear atom density of the cloud \\
$g = 2 \omega_r a_s$ & 1D interaction strength \\
$\mu = g n_0$ & chemical potential \\
$\omega_l = 2\pi \cdot 10 \unit{s^{-1}}$ & longitudinal trapping frequency in case of a harmonic confinement\\
$L = 200 \unit{\mu m}$ & total length of the simulation region \\
$\mathfrak{N}_z = 512$ & number of spatial discretization points, so the state $\psi(z)$ has $\mathfrak{N}_z$ degrees of freedom \\
$N_{\max} = 10^5$ & total number of Metropolis-Hastings iterations \\
$n(z) = |\psi(z)|^2$ & local density \\
$\phi(z) = \arg\,\psi(z)$ & local phase \\
\bottomrule
\end{tabular}
\label{t:2}
\end{table}

Typical examples of the grand canonical thermal state of the 1D Bose-Einstein quasicondensate after $N_{\max} = 10^5$ Metropolis-Hastings iterations are presented in figure~\ref{fig:profiles}.

The initial state for all the presented results was taken to be the ground state of the non-interacting gas ($n_0(z) = n_0 = \const$, $\phi_0(z) = 0$) in the untrapped case, and a Thomas-Fermi parabolic density profile with constant zero phase in the case of the harmonic confinement.

\begin{figure}
 \centering
 \begin{tabular}{cc}
  \includegraphics[width=0.46\textwidth]{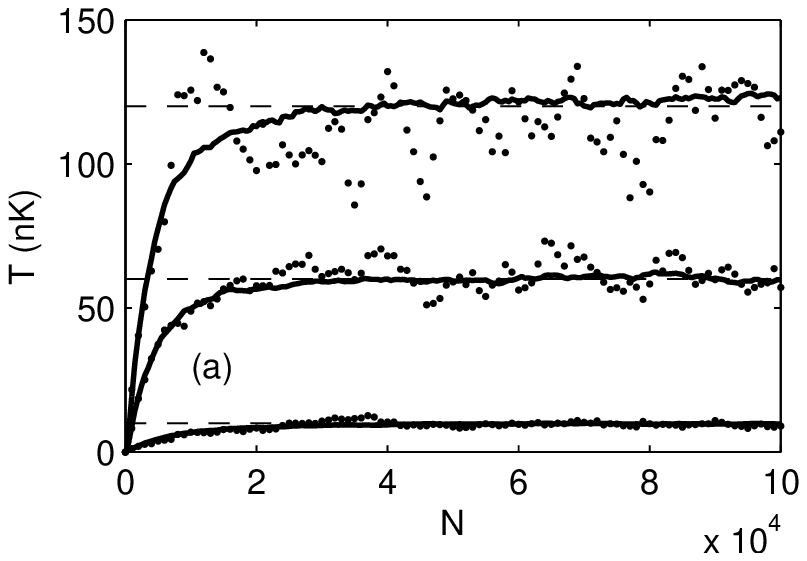}
    & 
  \includegraphics[width=0.46\textwidth]{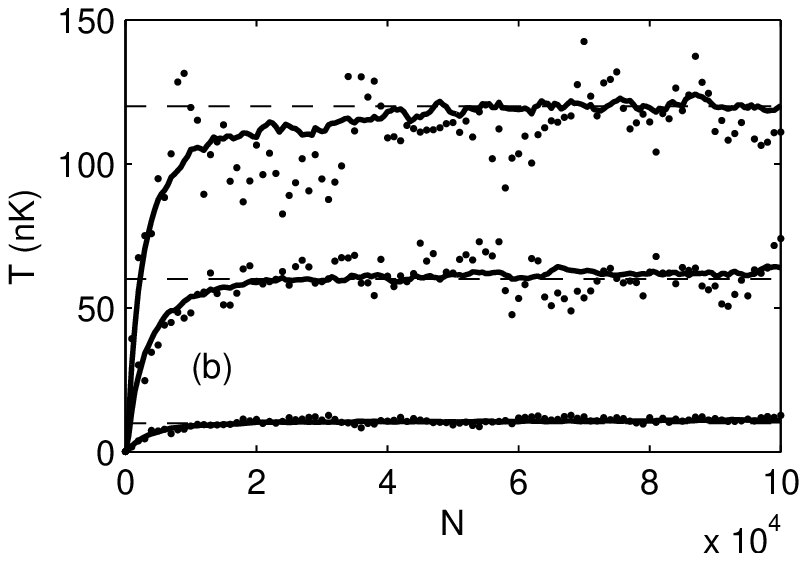}
 \end{tabular} 
 \caption{Temperatures during the Metropolis-Hastings `evolution' as a function of the iteration number $N$ in the case of untrapped (a) and harmonically trapped (b) gas for three equilibrium temperatures (given as external parameters) $T = 10, 60$ and $120$ nK. These temperatures are represented by three horizontal dashed lines serving as guides for the eye. Dots stand for one particular realization of the algorithm for the three temperatures (respectively, from bottom to top), and the corresponding solid lines show the averaged temperature over an ensemble of 70 realizations, each having the same initial conditions. Large temperature fluctuations in a single realization stem from the finite size of the simulation region, as they should converge to the equilibrium value only in thermodynamic limit. But from the ensemble averages it is evident that the thermal equilibrium is achieved after $N = 2-6 \cdot 10^4$ iterations.}
 \label{fig:temperatures}
\end{figure}

\begin{figure}
 \centering
 \begin{tabular}{cc}
	\includegraphics[width=0.46\textwidth]{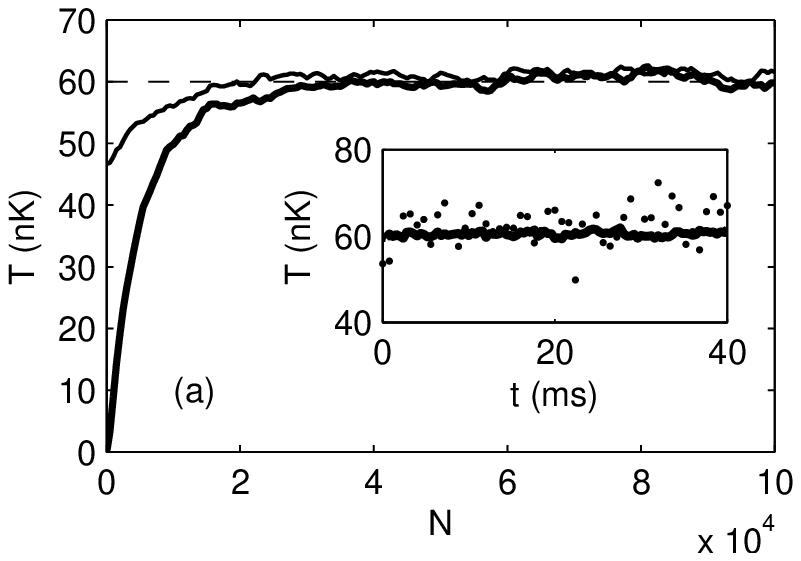}
    & 
  \includegraphics[width=0.46\textwidth]{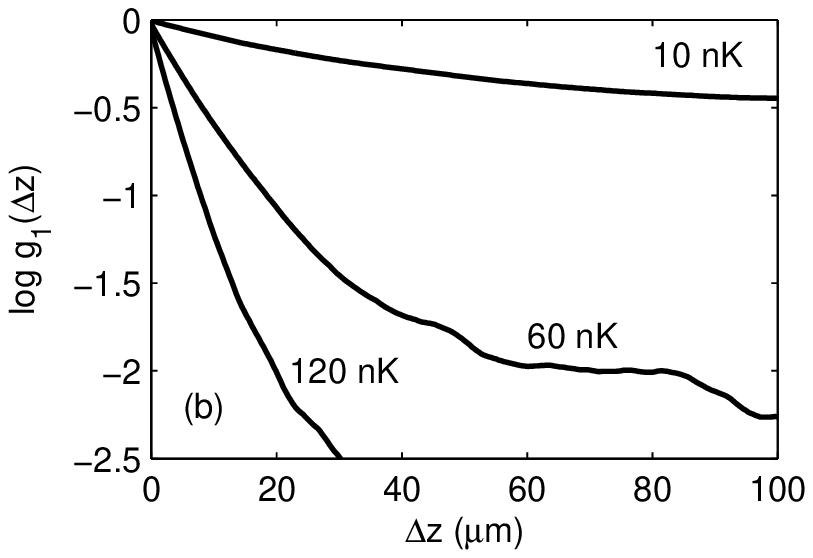}
 \end{tabular} 
 \caption{\textbf{(a)} Influence of the initial state on the rate of convergence to the thermal state. Temperatures during the Metropolis-Hastings `evolution' as a function of the iteration number $N$ in the case of untrapped  gas for $T = 60$ nK, averaged over 70 realizations. Thick line: zero-temperature state of the non-interacting gas, cf. figure~\ref{fig:temperatures}(a). Thin line: thermal gas of Bogoliubov quasiparticles with random phases and constant amplitudes (see explanation in the text). Both choices of initial conditions eventually lead to equilibrium, but in case of the `Bogoliubov gas' the convergence is faster, meaning that it is a better `initial guess' for the thermal state.
\textbf{Inset.} Temperature of the state, produced by the real-time GPE evolution starting from the achieved thermal state as a function of time. Dots: one single realization, solid line: average over 70 realizations. The stability of the temperature shows that the initial state was indeed the thermal state of the Gross-Pitaevskii Hamiltonian. 
\textbf{(b)} Natural logarithm of the $g_1$ correlation function in the homogeneous case for the temperatures $T = 10, 60$ and $120$ nK (from top to bottom) at the last iteration $N_{\max}$ of the algorithm, averaged over the ensemble of 70 realizations. These $g_1$ functions were used to calculate averaged temperatures presented in figure~\ref{fig:temperatures}(a). The linear region of the logarithm spans from 0 till $\approx 15\,\mu$m, and it is used in temperature measurement. The bending and fluctuations in the subsequent region are due to the finite size effects (as the total size of the system is $L=200\,\mu$m) and are to be discarded.}
 \label{fig:g1}
\end{figure}

The achievement of the steady state is controlled by temperature measurement at the each iteration of the algorithm, calculated from the $g_1$ autocorrelation function
\begin{equation}
g_{1,N}(\Delta z) = \frac{\int \psi_N^*(z)\,\psi_N(z+\Delta z)\,\rmd z}{\int |\psi_N(z)|^2 \rmd z}.
\label{e:g1}
\end{equation}

In thermodynamic equilibrium at positive temperatures in 1D $g_1$ is exponentially decaying with $\Delta z$, confirming the fact that there can be no true Bose-Einstein condensate in this case
\begin{equation}
g_1(\Delta z) = \rme^{-\Delta z/\lambda_T},
\end{equation}
where $\lambda_T$ is thermal coherence length
\begin{equation}
\lambda_T = \frac{2\hbar^2 \bar n}{m k_B T},
\end{equation}
with $k_B$ being the Boltzman's constant and $\bar n = \frac{1}{L'}\int |\psi(z)|^2 \rmd z$ the mean density of the cloud. $L'$ is the averaging length, which is the length of the integration region in Equation~\ref{e:g1} as well. In case of untrapped gas gas $L' = L$ is the total simulation region, but in case of harmonic confinement the integration region 
contains only the points where the local density $n(z)$ is larger than one tenth of the mean density. This helps to get rid of unessential boundary perturbations, probing the temperature of `the bulk' of the condensate.

The Metropolis-Hastings `evolution' of the temperature is presented in figure~\ref{fig:temperatures}, with one particular example of the $g_1$ function in figure~\ref{fig:g1}(b). It is evident that the thermal equilibrium is achieved after $N = (2-6) \cdot 10^4$ iterations.

Another independent test whether the achieved state is thermal is the real-time development of the state using Gross-Pitaevskii equation, as by definition the thermal state should remain thermal during such evolution. The results for the untrapped gas, presented in the inset to the figure~\ref{fig:g1}(a), show that indeed the temperature of the state does not change on average, assuring that the initial state was thermal with respect to the Gross-Pitaevskii Hamiltonian (there exist efficient algorithms for solving real-time GPE, see e.g. \cite{Vudragovic2012,Muruganandam2009}).

As in all realization of Metropolis-Hastings algorithm a `good guess' of the initial state is essential for the fast convergence. In figure~\ref{fig:g1}(a) we compare the beforementioned zero-temperature initial conditions with the initial state given by the thermal gas of Bogoliubov quasiparticles with random phases and constant amplitudes, given by the equilibrium Bose-Einstein distribution \cite{Grisins2011,Stimming2011}. This initial condition seems to be a much better `initial guess', leading to faster convergence. Note that the analytical method is only an approximation (implying weak interactions and neglecting the variance of the amplitudes of the quasiparticles) and doesn't immediately lead to the desired thermal equilibrium. This is one particular example where numerical methods successfully compete with the analytical ones.

\begin{figure}
 \centering
 \begin{tabular}{cc}
  \includegraphics[width=0.46\textwidth]{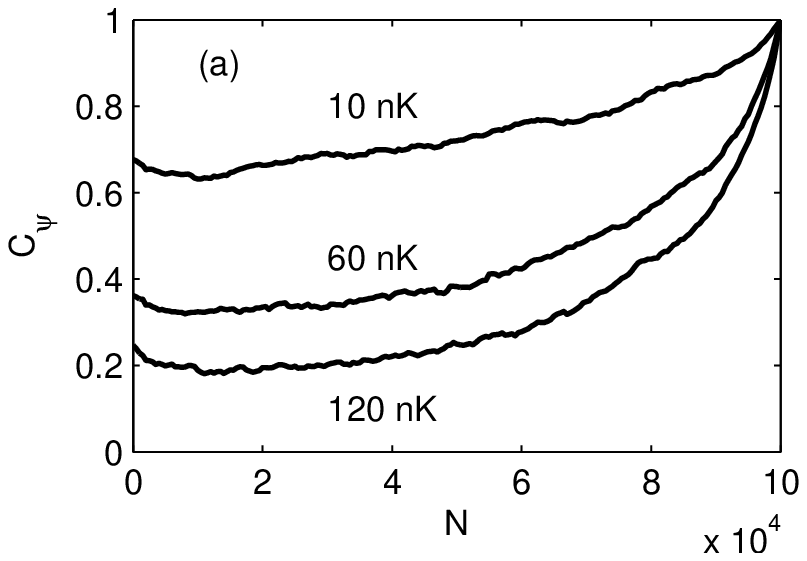}
    & 
  \includegraphics[width=0.46\textwidth]{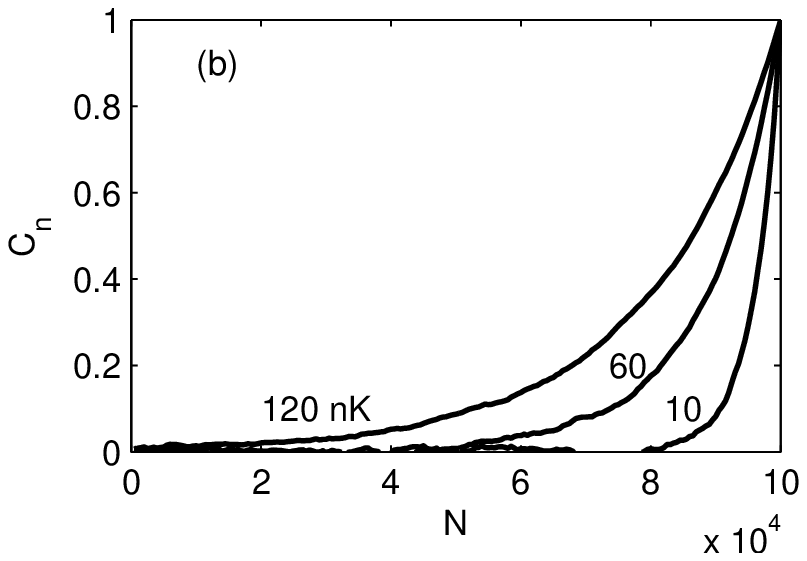}
 \end{tabular} 
 \caption{\textbf{(a)} Order parameter correlation function $C_\psi$ for the untrapped gas as a function of the iteration number $N$ for temperatures 10, 60 and 120 nK (from top to bottom), averaged over 70 realizations. Remaining strong phase coherence after $10^5$ iterations is due to the existence of long-range order in finite-size quasi-BEC. \textbf{(b)} Density fluctuation correlation function $C_n$ for the same realizations as in subfigure (a) for temperatures 10, 60 and 120 nK (from bottom to top).}
 \label{fig:correlations}
\end{figure}

It is well known that Markov chain methods give highly correlated samples from one iteration to the other. We present some correlators for the untrapped gas in figure~\ref{fig:correlations}, where $C_\psi$ is the two-point correlation function of the last sample $\psi_{N_{\max}}(z)$
\begin{equation}
C_\psi = \mathrm{Re}\;\frac{\int \psi_N^*(z)\,\psi_{N_{\max}}(z)\,\rmd z}{\sqrt{\int |\psi_N(z)|^2 \rmd z}\cdot \sqrt{\int |\psi_{N_{\max}}(z)|^2 \rmd z}},
\end{equation}
and $C_n$ is the density fluctuation correlation function of the last sample
\begin{equation}
C_n = \frac{\int \delta n_N(z)\,\delta n_{N_{\max}}(z)\,\rmd z}{\sqrt{\int \delta n_N(z)^2 \rmd z}\cdot \sqrt{\int \delta n_{N_{\max}}(z)^2 \rmd z}},
\end{equation}
where $\delta n(z) = n(z) - \bar n$, $n(z) = |\psi(z)|^2$, and $\bar n = \frac{1}{L'}\int\,n(z)\,\rmd z$ for the uniform gas.

It is evident that the order parameters still remain phase-correlated after $10^5$ iterations, which is consequence of the fact that we observe the system below the thermal gas to quasicondensate crossover temperature \cite{Bouchoule2007}: the thermal fluctuations are too weak to randomize the overall phase (note that the effects of phase diffusion are absent as there is no real-time propagation).

This remaining phase correlation has to be taken into account when performing simulations involving two or more independently prepared condensates, where a random constant overall phase difference should be added to the initial conditions at each realization. For one condensate it is not necessary, as only the phase difference is observable, and not the phase itself.

Density fluctuation correlation function $C_n$ gives a better representation of the correlations in Metropolis-Hastings algorithm, and from the numerical simulations it follows that one should pick one state out of $N_c = (2-8)\cdot 10^4$ iterations (depending on the temperature) to assure statistical independence. It is always a safe choice to pick only one last realization out of the whole Markov chain, reinitializing the simulation for each `measurement'. 

\section{Conclusion}

We developed an application of Metropolis-Hastings algorithm to sampling the classical thermal states of one-dimensional Bose-Einstein quasicondensates in classical field approximation in the case of untrapped gas with periodic boundary conditions and in experimentally relevant case of harmonic confinement. The achieved thermal steady state can be further used as an initial state for truncated Wigner simulations. In case when the quantum noise is important (e.g. collisions of condensates \cite{Perrin2007}, prethermalization of a split quasicondensate \cite{Gring2012}), it can be added to the thermal state using conventional methods \cite{Blakie2008,Polkovnikov2010}.

The proposed algorithm can be generalized to higher dimensions and arbitrary trap geometry. We see the main advantage of the proposed method in its ability to sample not only the conventional thermodynamic ensembles, but also the generalized Gibbs ensemble, which is believed to arise in the integrable one-dimensional bosonic gas \cite{Rigol2007,Kollar2011}.

\section*{Acknowledgements}
The authors are grateful for fruitful discussions with Wolfgang Rohringer, Bernhard Rauer, Tarik Berrada and J\"org Schmiedmayer. We acknowledge support from the FWF projects P22590-N16, Z118-N16 and Vienna Doctoral Program on Complex Quantum Systems (CoQuS).

\bibliography{../../Metropolis}

\end{document}